# Known allosteric proteins have central roles in genetic disease

György Abrusán [1*], David B. Ascher [2,3,4], and Michael Inouye [1,5-9]

[1] Cambridge Baker Systems Genomics Initiative, Department of Public Health and Primary Care, School of Medicine, University of Cambridge, UK

[2] Department of Biochemistry, University of Cambridge, Cambridge, UK

[3] Structural Biology and Bioinformatics, Department of Biochemistry, Bio21 Institute, University of Melbourne, Melbourne, Australia

[4] Computational Biology and Clinical Informatics, Baker Heart and Diabetes Institute, Melbourne, Australia

[5] Cambridge Baker Systems Genomics Initiative, Baker Heart and Diabetes Institute, Melbourne, Australia

[6] British Heart Foundation Cardiovascular Epidemiology Unit, Department of Public Health and Primary Care, University of Cambridge, Cambridge, UK

[7] British Heart Foundation Centre of Research Excellence, University of Cambridge, Cambridge, UK

[8] Health Data Research UK Cambridge, Wellcome Genome Campus and University of Cambridge, Cambridge, UK

[9] The Alan Turing Institute, London, UK

*Author for correspondence: György Abrusán,
email: gaa27@medschl.cam.ac.uk

**ABSTRACT**
Allostery is a form of protein regulation, where ligands that bind sites located apart from the active site can modify the activity of the protein. The molecular mechanisms of allostery have been extensively studied, because allosteric sites are less conserved than active sites, and drugs targeting them are more specific than drugs binding the active sites. Here we quantify the importance of allostery in genetic disease. We show that 1) known allosteric proteins are central in disease networks, and contribute to genetic disease and comorbidities much more than non-allosteric proteins, in many major disease types like hematopoietic diseases, cardiovascular diseases, cancers, diabetes, or diseases of the central nervous system. 2) variants from cancer genome-wide association studies are enriched near allosteric proteins, indicating their importance to polygenic traits; and 3) the importance of allosteric proteins in disease is due, at least partly, to their central positions in protein-protein interaction networks, and probably not due to their dynamical properties.

## INTRODUCTION

Allostery is a form of regulation of protein activity, that enables the cell to fine tune the spatial and temporal activity of certain proteins[1]. Allosteric proteins are typically characterised by more than one ligand-binding site: an orthosteric site, which is the binding site responsible for the main biological function of the protein, like catalysis; and one or more allosteric sites, which can be seen as "switches" that turn the activity of the protein on or off, by inducing conformational changes [1–4]. The actual mechanisms of allosteric signal transduction can be diverse, but in most proteins it can be explained by shifts between populations of different conformations, that are modified or induced by ligands binding at critical sites of the protein[5–7]. Allostery and the identification of allosteric binding sites have received considerable attention from the drug discovery community[8] for several reasons. First, some of the most frequent drug targets like GPCRs or protein kinases are frequently allosteric[9,10]. Second, the orthosteric binding sites of proteins are frequently highly conserved, which makes it difficult to design drugs that target these sites and at the same time are specific, without significant off-target effects. However, allosteric sites are much less conserved [11], and their high structural diversity offers the possibility of designing drugs that target allosteric proteins in a more specific way.

The fraction of proteins regulated by allostery while performing their normal biological function is unclear; kinases or GPCRs represent only a minority, and the number of known allosteric proteins[12] probably significantly underestimate their real number in the human proteome. Typically, allosteric proteins form protein complexes (see[13] for an overview and citations therein), and are particularly common in complexes where ligands connect several proteins in the complex[13] (with so called "polydesmic" ligands[14]). Most research on allostery has focused on the analysis of individual proteins or protein complexes, typically to identify the mechanism of allostery and allosteric pathways within them, or to find novel druggable allosteric sites. However, proteins perform their functions in the networked environment of the cell, e.g. metabolic networks, signalling networks, gene regulatory and protein interaction networks amongst others. Nussinov and colleagues suggested that allostery should be analysed at the level of cellular networks [15,16], because many key proteins in cellular signalling pathways (e.g. receptors) are allosteric [17–19], as allostery is well suited to propagate signals. However, a global, systems-level analysis of allostery and its role in genetic disease is currently missing, although a large-scale investigation of somatic cancer mutations found that they are enriched near allosteric sites [20].

In this work, we examine the role of allosteric proteins in genetic disease. We show that known allosteric proteins are much more likely to cause genetic disease than non-allosteric proteins, even if proteins involved in signaling (including kinases and GPCRs) are excluded from the data, and are central in disease networks. Our analysis shows that allosteric proteins contribute to hundreds of diseases, and are most common in those of the hematopoietic system, (cardio)vascular diseases, and cancers. The analysis of cancer GWAS data indicates that their variants are also enriched near allosteric proteins, indicating that their contribution is enriched for polygenic traits. Surprisingly, we find that the central role of allosteric proteins in disease is unlikely to be caused by their dynamical properties directly, but, at least partly, by their central positions in protein-protein interaction networks. This suggests that evolutionary pressure for fine-tuned regulation of network hubs is likely to contribute to the evolution of allostery, and that allostery is not so much the cause, but rather the consequence of the centrality of these proteins in disease and PPI networks.

**RESULTS**

**Data summary.** We compiled a list of 6170 proteins which are associated with genetic disease, i.e. harbouring pathogenic mutations from ClinVar[21], OMIM[22], Uniprot and HGMD[23] (Methods). As the disease annotations and names in different databases are not identical, we used the human disease ontology (doid.obo file, Methods), to standardise disease nomenclature and excluded proteins that could not be mapped to a disease term present in the disease ontology. In total, there were 5050 proteins associated with at least one disease ontology term which were available for downstream analyses. We used the Allosteric Database[12] (v4.10) to obtain known allosteric proteins. Of the 835 allosteric proteins in humans, 450 were associated with disease, and 380 are associated with at least one disease term present in the disease ontology.

**Allosteric proteins are enriched in many major disease types, including hematopoietic and vascular diseases, and cancers.** We performed disease ontology and gene ontology analyses, to examine whether there are diseases where mutations in allosteric proteins are particularly common. We identified the full list of disease ontology terms that map to each protein (diseases and their parental terms), and performed an ontology analysis similar to a gene ontology (GO) enrichment analysis. We found that allosteric proteins contribute to a large number of genetic diseases and conditions (Table S1): they are significantly enriched in 214 disease ontology terms (Table S1, $p < 0.05$, FDR), which include many of the major disease types: cardiovascular diseases, cancers, diabetes, rheumatoid arthritis and central nervous diseases (Figure 1A). The diseases where allosteric proteins are the most significantly enriched ($p < 0.005$ after Bonferroni correction, Figure 1B) are hematopoietic diseases (DOID:74) a diverse set of cancers, and vascular disease (DOID:178). Proteins associated with these diseases are characterised by a 2- to 5-fold enrichment among allosteric proteins compared to all disease proteins (Table S1).

To identify which functions of allosteric proteins contribute most to cancers, hematopoietic and vascular diseases, we performed a Molecular Function (MF) GO enrichment analysis on the proteins used in disease ontology analyses, and next a gene overlap analysis (see Methods) to test for overlaps between the lists of allosteric proteins of the significantly enriched disease ontology and MF GO terms. MF GO terms where allosteric proteins were overrepresented were not dominated by a few categories but have diverse functions (Figure 1C, $p < 0.05$ after Bonferroni correction, see Table S2 for full results and exact p-values). The gene overlap analysis identified several disease and MF gene otology terms with a significant overlap between cancer related terms and MF GO terms of protein kinases (Figure S1, $p < 0.05$ after Bonferroni correction), consistent with the known roles of kinases in carcinogenesis[24,25]. However, in the case of allosteric proteins of hematopoietic and vascular diseases no comparably strong overlap with proteins of MF GO terms was observed.

**Allosteric proteins cause more diseases than non-allosteric ones, and are central in disease-protein networks.** Next, we examined whether allosteric proteins were more likely to be involved in diseases than non-allosteric ones. As the frequency of allostery and the topology of allosteric pathways can vary substantially among different protein complexes[13], we performed both a pooled analysis of all allosteric proteins and assessed whether the patterns depend on the quaternary structure of proteins. Quaternary structure (i.e. heteromer, homomer, monomer) was assigned using the structures of the Protein Data Bank (PDB), and could be assigned to only 53% of proteins with disease annotations. The diseases associated with each protein were determined as previously, using the disease ontology terms, with the difference that the parental terms of diseases were not used, only disease terms that are independent from each other (see Methods). The analysis shows that allosteric proteins are more likely (2-fold, when all proteins are included) to cause disease than non-allosteric proteins (Figure 2A), and were also associated with significantly more diseases per protein (Figure 2B). This pattern is somewhat more pronounced for cancers than for non-cancers, but they were not qualitatively different (Figure S2). We found the number of diseases associated with allosteric vs non-allosteric proteins to be different also when stratified by quaternary structure; it is strongest in heteromers, and not significant in homomers (Figure 2B).

To examine the importance of allostery at the systems level, we constructed a disease-protein network, using the approach of Goh et al.[26] (see Methods). Each protein is a node in the network, nodes are connected with an edge if both proteins are associated with the same disease, and the weight of the edge is defined by the number of diseases. The resulting network (Figure 2F) has one large connected component with 3553 proteins, several small clusters, and 1068 isolated nodes. The analysis of the parameters of the nodes indicates that allosteric proteins had significantly higher betweenness centrality than non-allosteric proteins (except for proteins that form homomers, Figure 2C). Betweenness centrality measures the number of shortest paths in the network that pass through a particular node, and nodes with high betweenness centrality typically connect clusters and control most of the information flow in a network[27]. However, the pattern does not change qualitatively when all kinases and GPCRs, or all proteins involved in signal transduction are excluded from the analysis (Figure S3). Allosteric binding sites are frequently identified during screening for novel drug binding sites, raising the possibility that the central role of allosteric proteins in the disease network is due to being a drug target rather than due to being allosteric. Thus, we examined whether a similar pattern is present within pharmacologically active drug targets. We found that allosteric proteins have significantly higher betweenness than non-allosteric proteins, both when only drug targets (Figure 2D) or only non-drug targets were included (Figure 2E), indicating that being a drug target is not sufficient to explain the observed pattern, even though drug targets are characterised with higher betweenness.

**Allosteric proteins are involved in more comorbidities than non-allosteric proteins.** It has been demonstrated that genes involved in multiple diseases are also more likely to be responsible for comorbidities[28]. In addition, as clusters in disease networks usually correspond to proteins/genes of specific diseases[26], the nodes/proteins connecting them (i.e. with high betweenness centrality) are particularly good candidates for causing disease comorbidities[29]. To examine this, we analysed a recently published comorbity network, that was constructed using 17 million cases of the FDA Adverse Event Reporting System[30], without the use of genetic information. For proteins involved in at least two diseases, a significantly higher fraction of allosteric proteins was associated with comorbid diseases or conditions (Figure S4A), and allosteric proteins are also involved in significantly more comorbidities (Figure S4B). However, the relationship between the number of diseases and number of comorbidities between them is not qualitatively different for the two protein types (Figure S4C), indicating that the higher number of comorbid conditions of allosteric proteins is primarily the consequence of the higher number of diseases they are involved in.

**Common variants for cancers are enriched near allosteric proteins.** The genetic architecture of diseases is typically grouped into two broad categories: Mendelian diseases caused by a single or small number of low frequency genetic variants with strong phenotypic effects; and complex (polygenic) diseases caused by a large number of common genetic variants with weak effects. (The two types are part of a continuum though[31].) The ClinVar and HGMD databases are focused on Mendelian diseases. To examine whether allosteric proteins also have a preferential contribution to complex diseases, we asked whether genome-wide association studies (GWAS) are more likely to identify loci near allosteric proteins than expected by chance. As our previous analyses indicated a consistent enrichment of allosteric proteins in diverse cancers (Figure 1), and a large number GWAS have been performed for multiple cancer types[32], we focused our analysis on GWAS of cancers.

To obtain a list of variants associated with cancers in GWAS, we used the GWAS Catalog[33] (v1.0.2). Variants identified by GWAS are typically not the causal variants of the trait being analysed, but are correlated with the causal variants through linkage disequilibrium. Nevertheless, the gene closest to the index SNP at a locus is usually the causal gene[34]; thus, for each associated variant, we identified a candidate causal gene - the nearest protein-coding gene with at least one exon closer than 100kb to the variant, then examined the enrichment of allostery in the candidate causal genes. We chose 100kb as the cutoff distance, because most long-range transcription factors are located closer than 100kb to their target gene[35], however our results were similar when a 50kb threshold was used (not shown). In addition, for every base in the genome, we identified the closest protein coding gene (within 100kb), and we identified the total number of bases that are located closest to proteins from three groups: allosteric proteins, 'genetic disease' proteins (as defined above in the Data summary, i.e. 6170 proteins, excluding allosteric ones), or 'other' proteins (which do not belong to either of the prior two groups). The frequencies of the bases proximal to the three protein groups were used to normalize the frequencies of variants identified by GWAS (see Methods).

Cancer GWAS are highly variable in terms of sample size (and thus statistical power) and the number of identified variants per study, even for the same cancer types. The majority of studies include 1,000-10,000 cancer cases, and identify less than 10 significant ($p < 10^{-8}$) variants (Figure 3A and B). Cancers are characterised by high genetic heterogeneity even when they originate in the same tissues [36,37], and GWAS with large numbers of cancer cases are typically available only for the most common cancer types. We performed three different analyses which treat the underlying heterogeneity of the GWAS data differently. In the first analysis (Figure 3C-F), for every cancer type (mapped trait of GWAS catalog, see Methods) we used only the variants of the study with the highest number of cancer cases. In the second analysis (Figure 3G-J), we compiled a single nonredundant list of genes from all available studies of the same cancer type, and enrichment was calculated in the pooled lists. In the third analysis (Figure S5), we used the variants/genes of all studies which identified at least partially nonredundant lists of genes for every cancer type. In addition, in all three analyses, we used two different datasets: one that uses cancer GWAS studies irrespectively of the number of reported variants ("All"), and one that only use GWAS reporting less than 10 significant variants, which is less affected by the few commonest cancer types, and the included studies have less variability in statistical power to detect associations. The list of all GWAS used, along with their candidate causal genes is provided in Supplementary Table 4.

We found similar patterns of enrichment in all three analyses. Allosteric proteins show a significant, 2-fold enrichment compared to "other" proteins that are neither allosteric nor involved in disease (Figure 3C and G, Figure S5A), while 'genetic disease' proteins show a less pronounced, but still significant 1.5-fold enrichment (Figure 3C and G, Figure S5A). When using only the variants with the highest significance in each GWAS (20% with the lowest p-values), we found an even stronger, 3-fold enrichment near allosteric proteins compared to other proteins (Figure 3D and H, Figure S5B). In the analyses based on the studies reporting less than 10 variants, variants near allosteric proteins also show a high, 3-fold enrichment compared to other proteins of the genome, similarly to the most significant variants of the full dataset (Figure 3E and H, Figure S5C; note that a 3-fold enrichment means that 11.4% of GWAS variants is located close to allosteric proteins, while in the genome 3.8%). Taken together, these results indicate that, at least in cancers, genes of allosteric proteins are enriched as the nearest genes for common variants, and the effect is strongest among the genes near the variants with the highest significance.

**Allosteric proteins are enriched for pathogenic mutations.** The role of allosteric proteins in disease may be due to several factors. These include a generally greater functional importance of allosteric proteins, their distinct structural and dynamical characteristics, central positions in cellular networks, or research bias. To assess functional importance, we calculated the level of conservation for every disease associated protein using their mammalian orthologs, excluding primates (see Methods). Surprisingly, we found only a small, 2% difference between the conservation of allosteric (88.12%) and non-allosteric (85.99%) proteins (Figure 4A). We also performed the same comparison for kinases, to check whether allostery has an effect within a single protein family, and for drug target proteins, which are likely to be less variable in the research effort they receive. We found no significant difference in kinases (Figure 4B), but there is a significant effect (4% difference, $p = 7.47e-06$) in drug targets (Figure 4C).

Next, using ClinVar and HGMD mutations (see Methods) we examined whether there were consistent differences in the numbers of known pathogenic missense mutations of allosteric/non-allosteric proteins, and whether it influences the number of diseases they cause. Allosteric proteins have a significantly greater number of pathogenic mutations (2-3 fold) than non-allosteric ones, both when the entire dataset, kinases or drug targets are compared (Figure 4D,E and F). This may reflect true biological differences, e.g. allosteric proteins may be more vulnerable to mutation, but also could be caused by research biases, if allosteric proteins receive consistently more attention from the research community. To estimate bias, we examined whether the mutations of allosteric and non-allosteric proteins in ClinVar originate from same number of PubMed articles and found that there is a positive correlation between the number of diseases and number of articles reporting mutations (Figure S6). In addition, mutations of allosteric proteins are cited in significantly more research articles than non-allosteric ones (Figure S6). However, the greater number of articles per disease does not translate to a similarly greater number pathogenic mutations per disease, the relationship between the number of diseases and the number of pathogenic mutations is the same in allosteric and non-allosteric proteins in all three sets (Figure 4G, H and I). The difference is not significant for kinases and drug targets (Figure 4H and I), and significant but explain <1% of variance for the entire dataset (Figure 4G, ANCOVA was performed on log-transformed values). These results, in conjunction with those from GWAS indicate that the high importance of allosteric proteins in disease and disease networks is not simply the by-product of research bias. Furthermore, the positive correlation between the number of pathogenic mutations and diseases (Figure 4G-I) suggest that proteins associated with more diseases also have more vulnerabilities, possibly due to having more protein-protein interactions and interfaces.

**Pathogenic mutations accumulate in regions of 3D structures that are important in dynamics.** Allosteric proteins are characterised with multiple binding sites and allosteric pathways that connect them, thus the number of mutations that can interfere with the correct functioning of these proteins is greater than in the case of non-allosteric proteins, due to their dynamics. We examined whether the dynamic nature of these proteins contributes to their centrality in disease. First, we tested whether there are topological differences between the folds of conserved Pfam domains in allosteric and non-allosteric proteins. We mapped non-covalent residue interactions of the domains to a 50 x 50 matrix (Figure 5A and B, see also Methods), where position 1 is the N-terminus of the domain, and 50 is the C-terminus. We found that compared to non-allosteric proteins, allosteric proteins are depleted of long-range residue interactions, i.e. interactions connecting residues that are distant in the protein sequence (Figure 5C and D). This indicates that protein domains in allosteric proteins are more flexible than in non-allosteric proteins, as long-range interactions generally stabilise proteins and result in rigid folds.

Next, we examined whether disease associated mutations are enriched in the regions that are primarily responsible for the internal dynamics of proteins. Dynamic proteins can typically be partitioned into semi-rigid blocks of residues called "communities", which are characterised by correlated motions of residues [4,6,38]. The residues connecting the communities, particularly the ones characterised by high betweenness centrality in the residue-interaction network are the residues that are responsible for most of the motions of the proteins, including allosteric signal transduction (Figure 5J and K). We determined the community structure of the monomeric units of every disease-associated protein where a suitable PDB structure was available, using the STRESS tool[38,39] (Methods), and mapped the pathogenic mutations to the communities. We found a clear enrichment of pathogenic mutations in residues that connect communities, particularly the ones that take part in strong interactions, like H-bonds (Figure 5E and F). Surprisingly, the pattern is similar in allosteric and non-allosteric proteins, indicating that interference with the motions of proteins is important for pathogenesis in both cases.

Recently Kumar et al.[38] reported that cancer driver genes can be identified using communities, as somatic mutations of cancers are distributed unevenly across them, and driver genes are characterized by "hotspot communities" which are mutated much more frequently than others. We examined whether pathogenic mutations also show similar biases across communities, with respect to allostery. While the pathogenic mutations we use are not a mixture of drivers and passengers, their degree of pathogenicity is variable[40], and we expected that similarly to passengers in cancers, the less pathogenic ones will be distributed more evenly across communities. Our results show that pathogenic mutations are significantly more clustered than the random expectation (i.e. are present in fewer communities than expected, Figure 5G) and that the effect is stronger in allosteric proteins than in non-allosteric ones (p=9.37e-04, ANCOVA, Figure S7, and Figure 5G). This suggest that mutations in allosteric proteins are more pathogenic than in non-allosteric ones.

Allosteric signal transduction can cross protein-protein interfaces of complexes; therefore, we examined whether pathogenic mutations are distributed differently in the interfaces of known structures of allosteric and non-allosteric proteins. Our results indicate that, as it has been reported previously[41–43], pathogenic mutations are clearly enriched in interfaces (Figure 5H and I). The enrichment is more pronounced in heteromers (40-50%, Figure 5H) than in homomers (15-30%, Figure 5I), however the effect is not consistently stronger in allosteric proteins.

**The importance of allostery in disease networks is partly explained by protein-protein interaction (PPI) networks.** It has been shown that allosteric proteins have more connections in PPIs [20] than non-allosteric ones, thus the central role of allosteric proteins in disease may also be the result of their central positions in cellular networks. We examined this using two protein interaction datasets, IntAct, and the larger BioGrid. Using binary protein interactions, we constructed a PPI-network and determined the betweenness and degree centrality for every protein (Methods). The results show that, similar to the disease network, allosteric proteins in general, and particularly the ones forming heteromeric protein complexes, are characterised by significantly higher betweenness and degree centralities than non-allosteric disease proteins, both when IntAct (Figure S8A and B) or BioGrid (Figure S8C and D) were used to build the network. We found only a weak overall correlation between the centralities of the PPI and disease networks (Figure 6A, R = 0.063; and Figure 6B, R = 0.104), indicating that these two networks types have different topologies. However, allosteric proteins have consistently higher betweenness centralities than non-allosteric ones in both network types (Figure 6A and B), and are overrepresented among the proteins with the highest joint betweenness (both being above 1000, Figure 6C and D). This suggests that the more central position of allosteric proteins in PPI networks does contribute to the role of allostery in disease.

Hubs of PPI networks are frequently essential [27,44] (network centrality measures are even used to predict essentiality), therefore we examined whether the (at least monoallelic) loss-of-function in allosteric proteins is generally more deleterious than the loss-of-function for non-allosteric disease proteins. We used the distributions of premature stop codons (PSCs) of the gnomAD database[45] to determine the inactivation tolerance of every protein (see Methods). We found that allosteric proteins are characterised by a modest reduction in PSC density, and also by slightly lower mean allele frequencies of their PSCs compared to non-allosteric proteins (Figure S9). However, the fraction of proteins that are intolerant to inactivation, i.e., have no detected PTCs in their sequence (and are most likely haploinsufficient or essential) is similar in both groups (9.7% vs. 8.7%, p = 0.561, test of proportions). Taken together, these results indicate that inactivating allosteric proteins is somewhat more deleterious than inactivating non-allosteric disease proteins, but there are no significant differences in the essentiality of these two protein groups.

**DISCUSSION**

Our results indicate that the currently known allosteric proteins are much more important in genetic diseases than other proteins: they cause disease more frequently, and are associated with more diseases than non-allosteric proteins, primarily in cancers, hematopoietic and (cardio)vascular diseases (Figure 1 and 2). In addition, they are central in disease-protein networks, and are likely to be responsible for more disease comorbidities than non-allosteric proteins (Figure 2, Figure S3). Importantly, we also observed a clear enrichment of variants near allosteric proteins in cancer GWA studies (Figure 3), indicating that allosteric proteins also have higher than average contribution to complex disease.

The high importance of allosteric proteins in disease can be caused by at least two processes. First, as they have distinct dynamical properties[46] (i.e. signal transduction pathways that connect allosteric and orthosteric sites within the protein), they might have more residues that can make them inactive when mutated, thus they might simply be more vulnerable to mutations. The fact that allosteric proteins have more known pathogenic mutations and are somewhat more conserved (Figure 4) does support this. The analysis of the dynamical properties of the proteins in our dataset indicates that, as it has been suggested previously[46], mutations that alter dynamics are important contributors to disease, i.e. pathogenic mutations are overrepresented in residues important in transducing motions within proteins, and also in interfaces (Figure 5). However, except the more pronounced clustering of mutations in residue-communities, we found no dramatic differences between allosteric and non-allosteric proteins, suggesting that proteins involved in disease are generally dynamic, also in the non-allosteric group.

    A second possibility is that known allosteric proteins are characterised by central positions in cellular networks, that result in higher-than-average contribution to disease. Our finding, that allosteric proteins have high betweenness centralities both in disease and PPI networks do support this hypothesis (Figure 6, Figure S8), as well as their somewhat lower tolerance for loss-of-function (PSC) mutations (Figure S9). It has been argued that allostery is common among signalling proteins[16,17], and signalling pathways are a subset of protein-protein interaction networks. Since signalling proteins are frequent contributors to disease, we also examined whether involvement in signaling is alone sufficient to explain the observed patterns. We found that removing all proteins involved in signaling from the data (or all kinases and GPCRs) does not change our results qualitatively (Figure S3), therefore signaling alone cannot explain the pattern, and it is likely to be caused also by other or more general processes, for example evolutionary pressure for the emergence of fine-tuned regulation of nodes of high importance in cellular networks.

    The number of studies that investigate the mechanistic aspects of allostery is large, however only a few have examined the biological factors that are responsible for its evolution. It has been suggested that the origins of allostery are rooted in protein evolvability itself and not necessarily in function[47], and that potentially allosteric residues and pathways are "prewired" in proteins in the form of networks of coevolving residues[48,49], which are conditionally neutral, and emerge due to adaptations to fluctuating conditions[47]. However, recently it has been demonstrated that allostery emerges particularly frequently in proteins which have a large number of homologs with conserved binding sites in the genome[13] (like kinases), suggesting that a major evolutionary force behind the emergence of allostery is preventing cross-reactivity among them. The finding that allostery is a characteristic of nodes with high betweenness in disease and PPI networks indicates that network centrality, i.e., evolutionary pressure for regulating nodes critical in disease and PPIs is also a driving force behind it.

**METHODS**
**Data sources**. The list of known human allosteric proteins was downloaded from the Allosteric Database (v4.10), altogether 835 (from the total 1942). The list of proteins involved in disease was compiled from four sources: we used RefSeq transcripts from ClinVar[21] with mutations annotated as "pathogenic" or "likely pathogenic"; OMIM[22] genes; the list of disease proteins compiled by UniProt (https://www.uniprot.org/diseases/); and RefSeq transcripts from HGMD[23] (version 2014) with missense mutations annotated as "disease causing mutation". The list of pharmacologically active drug target proteins was downloaded from DrugBank (https://go.drugbank.com/).

**Ontology analyses.** The human disease ontology[50] file (doid.obo) was downloaded from The Open Biological and Biomedical Ontology Foundry

(http://www.obofoundry.org/ontology/doid.html). This ontology file was used to standardise disease nomenclature in all further analyses. For each protein of the full list of disease associated proteins, its MedGen identifiers, OMIM identifiers, and disease names were mapped to disease ontology IDs (DOIDs), provided by the doid.obo file. To reduce inconsistencies in disease names (HGMD does not provide MedGen or OMIM IDs, only disease names/descriptions), disease names were converted to lowercase, and commas were removed. Only those proteins were included in the ontology analyses where their MedGen/OMIM/disease names could be converted to a disease ontology ID, altogether 5050. Next, the full list of disease ontology IDs of every protein was determined by traversing the disease ontology hierarchy to the highest level "Disease" term, and a standard ID enrichment analysis (similar to a GO enrichment analysis) was performed with the GeneMerge tool[51]. In addition, we performed a GO enrichment analysis on the same set of proteins, using the mappings between UniProt proteins and GO Molecular Function terms, provided by the standard human gene ontology annotation file (goa_human.gaf), which was downloaded from http://geneontology.org/docs/downloads. The significance of the overlaps between the protein lists of enriched disease ontology and gene ontology terms was calculated using Fisher's exact test of the GeneOverlap R/Bioconductor package.

For visualisation and redundancy filtering of the significant molecular function GO terms, we used the REVIGO tool[52] with medium similarity setting. For disease ontology terms we used a methodology comparable to REVIGO, except that semantic similarities between the disease ontology terms were calculated with an edge-based method[53], which in the case of the disease ontology was more robust than node/information content based methods. We calculated the length of the shortest path between every possible pair of the enriched terms with igraph[54]. Next, we calculated a similarity metric for every pair using the path lengths:
$sim = (length_{max} - length_{ij} + 1) / length_{max}$ where $length_{ij}$ is the number of nodes on the shortest path between nodes i and j, and $length_{max}$ is the length of the longest path. We filtered out the most redundant nodes by keeping only the more significant one of the node pairs with sim > 0.9. (In the case of similar significances, the term located higher in the ontology was kept.) Finally, we used multidimensional scaling (cmdscale function of R) to reduce the dimensionality of the matrix of the similarities, and plot them a in a two-dimensional semantic space.

**Construction and visualization of disease-protein networks.** Disease-protein networks were constructed using an approach outlined by Goh et al.[26] For every protein we determined the number of diseases it contributes to, and a network was constructed where the nodes are proteins, which are connected with an edge if they are associated with the same disease. The number of diseases shared by the two proteins were used as the weight of the edge. The list of diseases associated with each protein was determined with a comparable, but more restrictive method than used for DOID enrichment analysis. Similarly, using MedGen IDs, OMIM IDs, and disease names we identified the list of disease ontology IDs associated with each gene. For diseases names a "bag of words" approach was used, i.e. disease names containing the same words were treated as similar, irrespectively of word order. This corrects for common inconsistencies in disease names like "myopathy, congenital" and "congenital myopathy", in cases where MedGen or OMIM IDs were not present. In addition, in cases where proteins were associated with multiple disease ontology IDs, we filtered out non-independent ones, leaving in only the IDs at the lower levels of the ontology, by removing higher level terms that are located on the same path towards the top-level "disease" (DOID:4) term of the ontology. For example, if a protein is associated with the terms "myeloid leukemia" and "leukemia", then only "myeloid leukemia" was used.

The betweenness centrality of the network nodes (proteins) was determined using igraph[54] (v0.8.3). The largest connected component of the network was visualized using the OpenOrd algorithm of the Gephi tool (v0.92), smaller components of the network with the DrL layout algorithm of igraph.

**Calculating the number of comorbidities for proteins.** We used only proteins that are involved in minimum two diseases (DOID terms, see above) in the disease-protein network. For every disease term we identified all their parental terms, up to the highest level "disease" term (DOID:4), using the disease ontology (doid.obo file). Next, we converted the disease ontology terms to MedGen IDs, using the mapping between DOID terms and MedGen/OMIM IDs of the disease ontology. OMIM IDs were converted to MedGen IDs using the MedGen/HPO/OMIM mapping file provided by MedGen (available at https://ftp.ncbi.nlm.nih.gov/pub/medgen/). The comorbidity network[30] based on FDA Adverse Event Reporting System (FAERS) was downloaded from http://nlp.case.edu/public/data/FAERS_comb/. The network contains 25217 edges between 1059 MedGen IDs describing diseases and conditions. The number of comorbidities for every protein was calculated by identifying every possible pairwise combination of their MedGen IDs, and counting the number of combinations that are present in the comorbidity network.

**Calculating enrichment of GWAS variants.** The GRCh38 release of the human genome, and the corresponding GTF file of genes were downloaded from Ensembl, release 101. Significant variants reported by all known cancer GWAS were downloaded from the GWAS catalog [33]; in each study only variants with $p < 10^{-8}$ were used. The type of cancer was assigned as the "mapped trait" column of the associations file, after filtering out studies with disease descriptions that were not explicitly focusing on cancer risk genes, for example "Toxicity response to radiotherapy in prostate cancer". Additionally, we excluded all studies using targeted genotyping arrays (Oncoarray), because they are not powered to be genome-wide. If two cancer GWAS had exactly the same variants, we used only one study in the analysis, but overlaps between the variant sets of two studies were permitted. We identified the closest protein coding exon for every known base of the genome, and for every variant in the GWA studies. Bases and variants that were further than 100kb from the closest protein coding exon were not assigned to any gene. If the variants of two GWA studies of the same cancer type were located near the same set of genes (i.e. identified the same sets of candidate causal genes, or a subset of the other), then only one study was used (with the higher number genes, or cancer cases). Ensembl genes were mapped to Uniprot proteins with the ID mapping tool of Uniprot. The enrichment of variants that are located closest to a particular protein type (e.g., allosteric) was calculated as the fraction of all variants of the pooled GWA studies, divided by the fraction of bases in the genome (% in GWAS / % in the genome). Pooling data from several GWAS allows for the meaningful analysis of GWAS with a few variants (and the comparison with larger ones), because in these studies a single variant can result in large changes in enrichment: a single allosteric variant in a GWAS with five variants means 5.26x enrichment compared to the genomic average. In the gene enrichment analysis, a single list of genes was compiled for every cancer type from all of its GWA studies, and the enrichment of allosteric, disease and other proteins was calculated compared to their frequencies in all protein coding genes of the genome. Significances of enrichment were calculated with the "metafor" R package using the heterogeneity test of the Mantel-Haenszel method (rma.mh).

**Calculating conservation.** We used the eggNOG (v5) orthology database[55] to calculate conservation. We downloaded the mammalian dataset, and calculated the level of conservation of human proteins using the multiple-alignments of the orthogroups. For every human protein in the alignment, we calculated their similarity with all non-primate proteins, as the fraction of the similar residues in the total number of aligned (i.e. non-gapped) residues in their pairwise alignment. Primates were excluded to ensure that all proteins in the alignment had approximately the same time to accumulate differences compared to their human ortholog, as most extant placental mammal groups diverged from primates in the mammalian radiation 70-80 mya[56]. The conservation of each human protein was defined as the average similarity with all other non-primate proteins of the alignment.

**Identification of conserved domains and residue interactions.** Raw conserved domains of disease associated proteins were identified with the hmmscan tool of HMMER 3.3[57], using Pfam-33.1 and an e-value cutoff (--domE) 0.001, and bitscore cutoff 22. When several domains mapped to the same region, we used only the most significant (typically the longest) hit. Next we mapped PDB structures to the identified Pfam domains with muscle[58][43], using structures with the highest resolution (and no worse than 3Å), and with the highest coverage, with a minimum of 90%; excluding domains shorter than 50 residues. Residue interactions within protein structures were calculated with the RINerator tool[59]. Residue interaction matrices were calculated as in[14] by dividing the length of each domain into 50 units. The interaction scores of each cell of the resulting 50 x 50 matrix were calculated by adding the number of residue-interactions of residue-pairs that fall into any given cell, using the *_nrint.ea files produced by RINerator. The number of residue-interactions was divided by the total number of interactions in the domain, to correct for differences in domain size. Additionally, the interaction scores of non-allosteric domains were normalized by the proportion of allosteric to non-allosteric proteins, to correct for the difference in the number Pfam domains in allosteric and non-allosteric proteins. In the final analyses Log transformed scores were used, to reduce the variation between cells.

**Analysis of dynamics.** For proteins associated with disease (allosteric and non-allosteric) we selected structures that cover at least 75% of the length of the protein, has the largest possible ligand (if it has a ligand), resolution is better than 3Å, and the structure has less than 12 chains. Altogether 563 proteins have such structures, 131 allosteric and 432 non-allosteric (see Supplementary Table 5). We used STRESS[39] to identify communities and critical residues in the structures. The MMTK-2.7.12 dependency of STRESS was installed with the install script by the lab of Pierre-Nicholas Roy at https://github.com/roygroup/mmtk_install. Before running STRESS the structures were preprocessed: if they were part of a protein complex than only the monomeric unit was used (a single chain with the disease protein), and we processed the structure with the DockPrep tool of Chimera[60], to add hydrogens, complete incomplete side chains, and remove residues with low occupancy when residues with alternative locations are present. In addition, similarly to [13] we modified STRESS to use the center-of-mass of residues instead of their C-α atoms, using the Bio3D R package [61] and in-house Perl scripts, as it was shown to significantly improve performance both in molecular dynamics[62] and elastic-network based simulations[13].

**Protein-protein interaction networks.** IntAct (10[th] Nov. 2020) and BioGrid (v.4.1.190) interaction datasets were downloaded from their corresponding websites. Only protein-protein interactions were used for both (annotated as "physical association" for IntAct and "physical" for BioGrid). Alltogether 139010 distinct interactions were identified in IntAct and 434534 in Biogrid; a weight 1 was assigned to all edges of the network. Betweenness and degree centralities of the network were calculated with igraph.

**Determining loss-of-function intolerance.** The gnomAD exome dataset[45] was downloaded from https://gnomad.broadinstitute.org/downloads. We discarded all variants from the dataset that did not pass all quality filters, or where the allele frequency was given as 0. For every Ensembl protein in the dataset we determined the number of PSCs that map to them (annotated as "stop_gained"), the length of the protein, and the average allele frequency of the PSCs per protein. When multiple SNPs in the same codon resulted in the same amino acid change (for example in the case of Tryptophan), the allele frequency of the amino acid change was determined as the sum of the allele frequencies of the variants. From the protein isoforms of the same genes, we used only the main isoform, as defined by the Appris database[63] (Gencode19/Ensembl74 dataset). Ensembl proteins were mapped to Uniprot proteins by the ID mapping service of Uniprot.

**ACKNOWLEDGEMENTS**
We thank Sergio Ruiz-Carmona and Loïc Lannelongue for critical reading of the manuscript. G.A. and M.I were supported by the Cambridge-Baker Systems Genomics Initiative. M.I. was supported by the Munz Chair of Cardiovascular Prediction and Prevention. This work was supported by core funding from the: British Heart Foundation (RG/13/13/30194; RG/18/13/33946), BHF Cambridge Centre of Research Excellence (RE/13/6/30180), and NIHR Cambridge Biomedical Research Centre (BRC-1215-20014)*. It was also supported by Health Data Research UK, which is funded by the UK Medical Research Council, Engineering and Physical Sciences Research Council, Economic and Social Research Council, Department of Health and Social Care (England), Chief Scientist Office of the Scottish Government Health and Social Care Directorates, Health and Social Care Research and Development Division (Welsh Government), Public Health Agency (Northern Ireland), British Heart Foundation and Wellcome. Additionally, this work was supported by the NHMRC GNT1174405 grant for D.A. and in part by the Victorian Government's Operational Infrastructure Support Program.
The authors declare no competing interests. *The views expressed are those of the author(s) and not necessarily those of the NIHR or the Department of Health and Social Care.

**AUTHOR CONTRIBUTIONS**
G.A. conceived the project, performed the analyses, and wrote the first version of the manuscript. D.A. contributed data, M.I. contributed to the writing of the manuscript.

**SUPPLEMENTARY MATERIALS**
Supplementary information of 9 figures and 5 tables are available with the online version of this paper.

**FIGURES**

**Figure 1.** Enrichment of allosteric proteins in disease ontologies and gene ontologies. **A)** Summary of the disease ontology terms where allosteric proteins are significantly enriched. The terms were filtered for redundancy and plotted using their semantic similarities, the ones highlighted were selected manually from the filtered list of terms. Allostery is overrepresented in a large number of diverse diseases, which include many of the major disease types like cardiovascular diseases, metabolic diseases (e.g. hypercholesterolemia, diabetes) central nervous system diseases, and cancers. **B)** Graph of the disease ontology terms where allosteric proteins are most significantly enriched. The intensity of red corresponds to significance, only terms with $p < 0.005$ (after Bonferroni-correction) are plotted. Allosteric proteins are most significantly associated with diseases of the hematopoietic system, cancers and vascular disease. **C)** GO analysis of significantly enriched Molecular Function terms. Related terms were grouped and visualized with REVIGO using their sematic similarity. The list of proteins in each term, and p-values are provided in Table S2. The analysis shows that allosteric proteins are involved in diverse functions, and the enriched terms are not dominated by a few categories.

**Figure 2.** Allosteric proteins are overrepresented in disease, and are central in disease-protein networks. **A)** Allosteric proteins are significantly more frequently involved in disease than non-allosteric proteins, irrespectively of their quaternary structure. * In the 'All' category the number of non-allosteric proteins was defined as human SwissProt proteins minus all (human) allosteric proteins. Note that the 'All' category includes also the proteins where quaternary structure is not known (i.e. have no entry in the PDB). **B)** Allosteric proteins (except homomers) are involved in significantly more diseases than non-allosteric ones. **C)** Allosteric proteins, particularly the ones forming heteromeric complexes have significantly higher betweenness centrality in the disease-network than non-allosteric ones. **D-E)** Drug-targets have generally much higher betweenness centralities than non-drug target proteins, however, allosteric proteins have significantly higher betweenness than non-allosteric proteins in both groups. **F)** The disease-protein network. Allosteric proteins are represented by red nodes, non-allosteric ones by blue, the size of the nodes was calculated as 1 + Log(nr of diseases). The largest connected component was visualized with the OpenOrd algorithm.

**Figure 3.** Variants of cancer GWAS are enriched near allosteric proteins. **A-B)** Relationship between the number of cancer cases and significant ($p < 10^{-8}$) variants in cancer GWA studies where at least one significant variant is located near a protein coding gene. The majority of GWAS have a moderate number of cases (and power), $10^3$-$10^4$, and identify less than 10 significant variants. Enrichment near different protein types was calculated in two datasets; one using studies irrespectively of the number of significant variants ("All"), and one using GWA studies reporting less than 10 significant variants, which is influenced less by the largest studies of the most common cancer types, and has less variability in statistical power to detect significant associations. **C-F)** In the analysis which only uses the studies with the highest number of cancer cases, allosteric and (mostly Mendelian) disease associated proteins show a 2- and 1.5-fold enrichment compared to other (i.e. neither allosteric nor disease associated) proteins (C). Variants with the highest significance in each GWAS (20% with the lowest p-values) show an even more pronounced, 3-fold enrichment near allosteric proteins (D). In studies reporting less than 10 variants, the enrichment near allosteric proteins is also 3-fold, comparable to the pattern seen with the most significant variants of the full dataset (E). The distribution of main cancer types in the variants of the two GWAS sets (F). **G-J)** The analysis which uses a single nonredundant list of genes compiled from all studies of the same cancer type (mapped trait of GWAS Catalog) shows a similar degree of enrichment of allosteric and disease proteins among the proteins identified by GWAS.

**Figure 4.** Conservation, mutation density, and disease associations of allosteric and non-allosteric proteins involved in disease. Kinases and drug targets are also shown separately. **A-C)** Allosteric proteins are characterised by a somewhat higher conservation than non-allosteric proteins. **D-F)** Allosteric proteins are characterised by significantly higher numbers of pathogenic mutations than non-allosteric ones. **G-I)** Despite having more pathogenic mutations, the relationship between the number of diseases, and the number of pathogenic mutations is qualitatively similar for allosteric and non-allosteric proteins.

**Figure 5.** Structural and dynamical characteristics of pathogenic mutations in allosteric and non-allosteric proteins. **A)** Residue interaction matrix of Pfam domains of allosteric proteins. **B)** Residue interaction matrix of Pfam domains of non-allosteric proteins. **C-D)** The difference between the two matrices (panel C) shows that allosteric proteins have significantly fewer long-range interactions in their Pfam domains than non-allosteric proteins (panel D), and are likely to be more flexible. **E-F)** Disease associated mutations are significantly enriched in community interfaces, i.e. residues that interact with members of other communities, both in the case of allosteric and non-allosteric proteins. The enrichment is more pronounced for stronger interactions like H-bonds (panel E). **G)** The distribution of disease mutations across communities (horizontal bar) differs from the random expectation (violin plot) more in allosteric proteins than in non-allosteric ones, indicating that pathogenic mutations have a stronger effect in allosteric proteins. See also Figure S7. **H)** Disease associated mutations are significantly enriched in the protein-protein interfaces of both allosteric and non-allosteric heteromers. **I)** A much less pronounced, but also significant enrichment is present in the interfaces of homomers. **J)** The structure of Mitogen-activated protein kinase 8 (MAPK8, PDB ID: 4qtd). **K)** Community structure of MAPK8. Each community is represented by a different colour, residue-residue interactions between different communities (community interfaces) are indicated with black.

**Figure 6.** Allosteric proteins have higher betweenness both in PPI and disease networks than non-allosteric proteins, nevertheless the overall correlation between the centralities of the two network types is weak. **A-B)** Correlations between PPI and disease protein network centralities. Proteins with 0 betweenness centrality in any of the two networks were excluded. **C-D)** Density plots of allosteric proteins. 43.5% (IntAct) and 39.9% (BioGrid) of proteins have betweenness centrality higher than 1000 in both networks (red rectangle). **E-F)** Density plots of non-allosteric proteins. 23.6% (IntAct) and 23.3% (BioGrid) of proteins have betweenness centrality higher than 1000 in both networks (red rectangle). Data ellipses on panel A and B were drawn with the stat_ellipse() function of ggplot2 (R), with default settings, ANCOVA was performed on log transformed data.

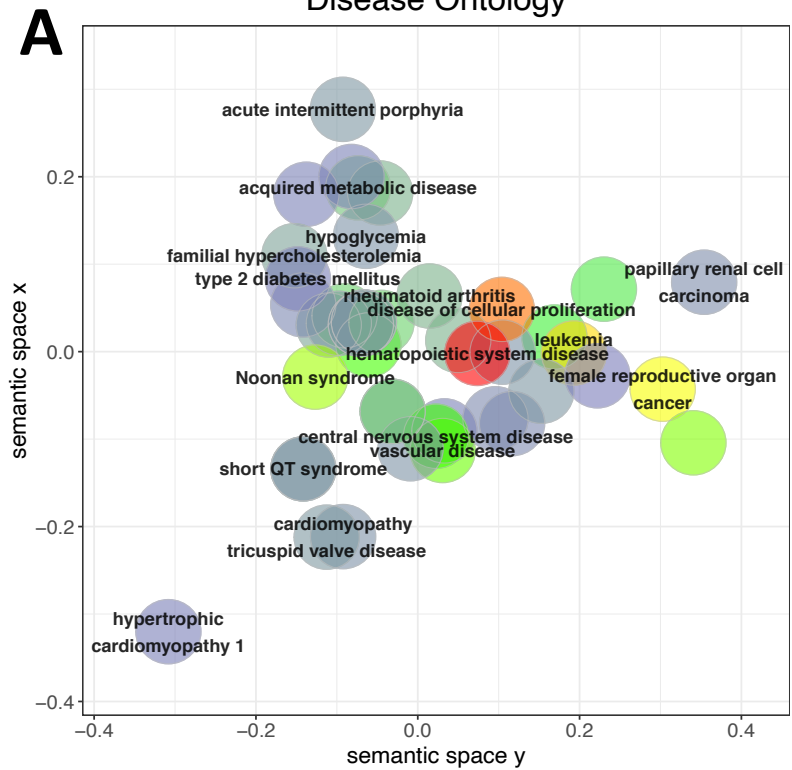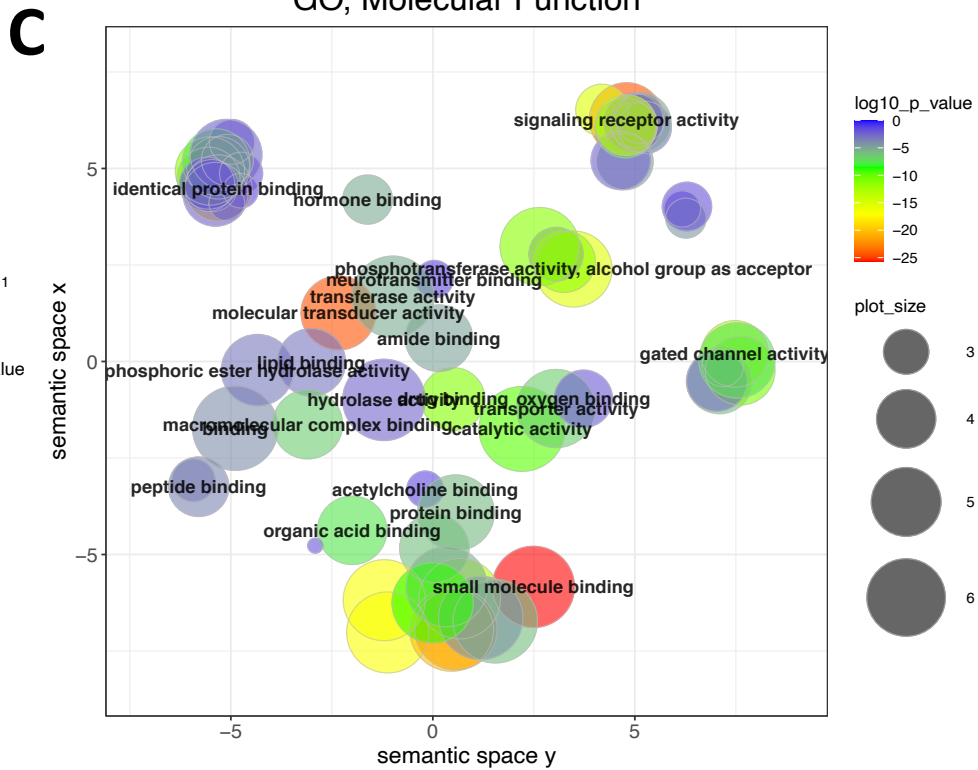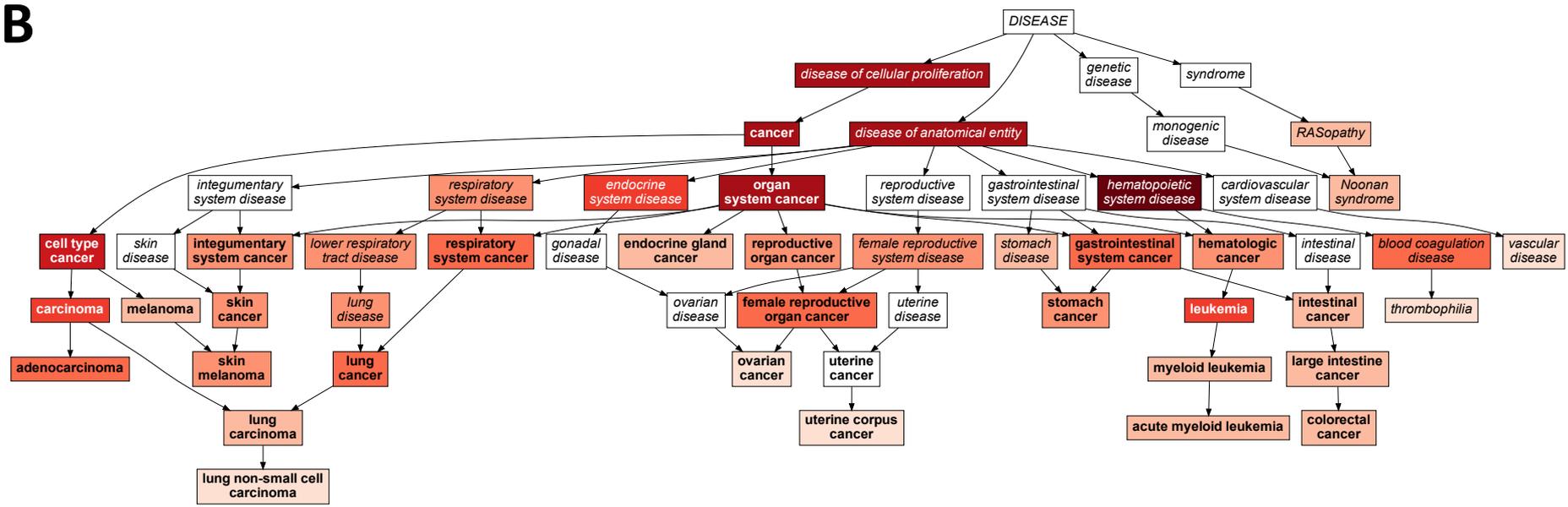

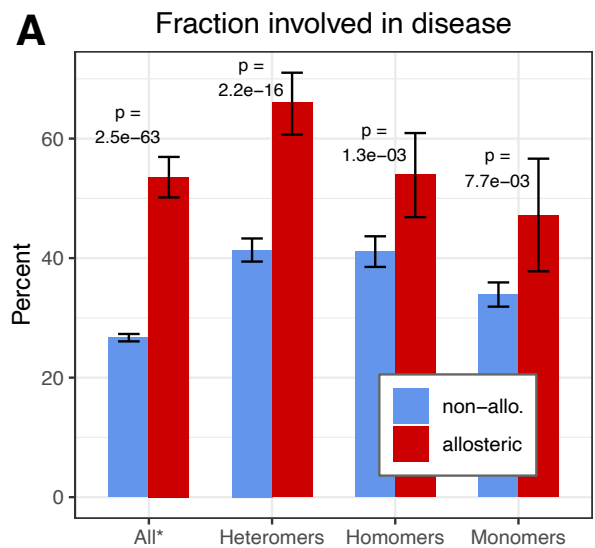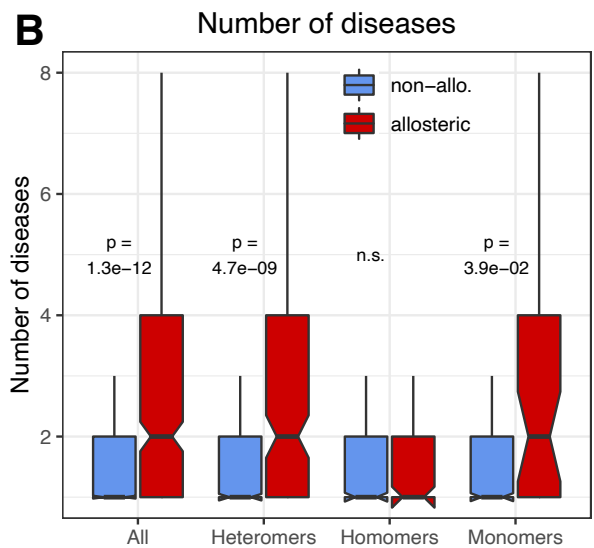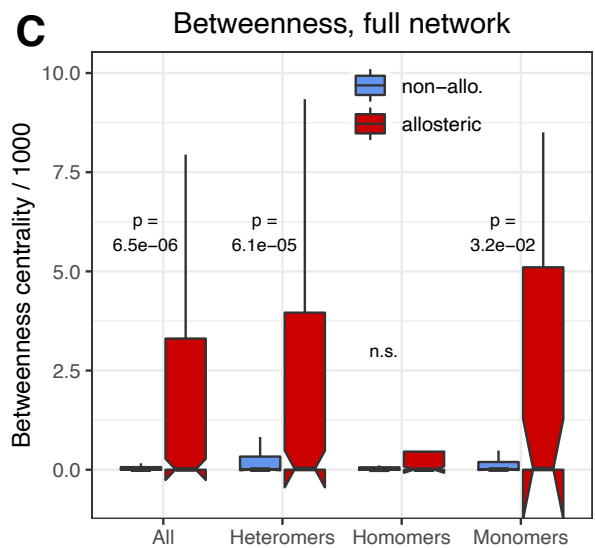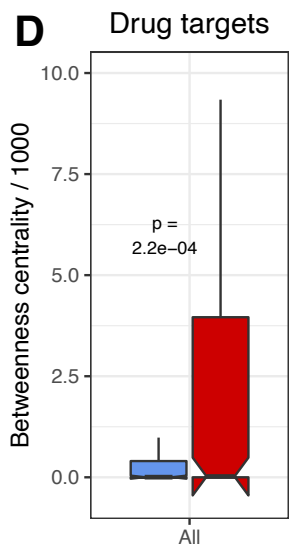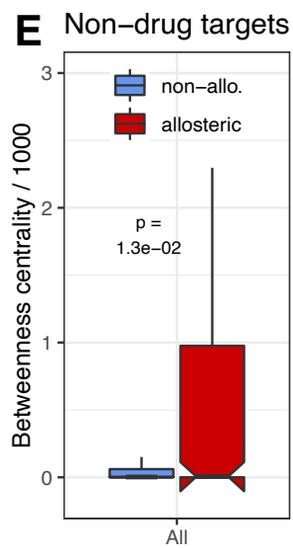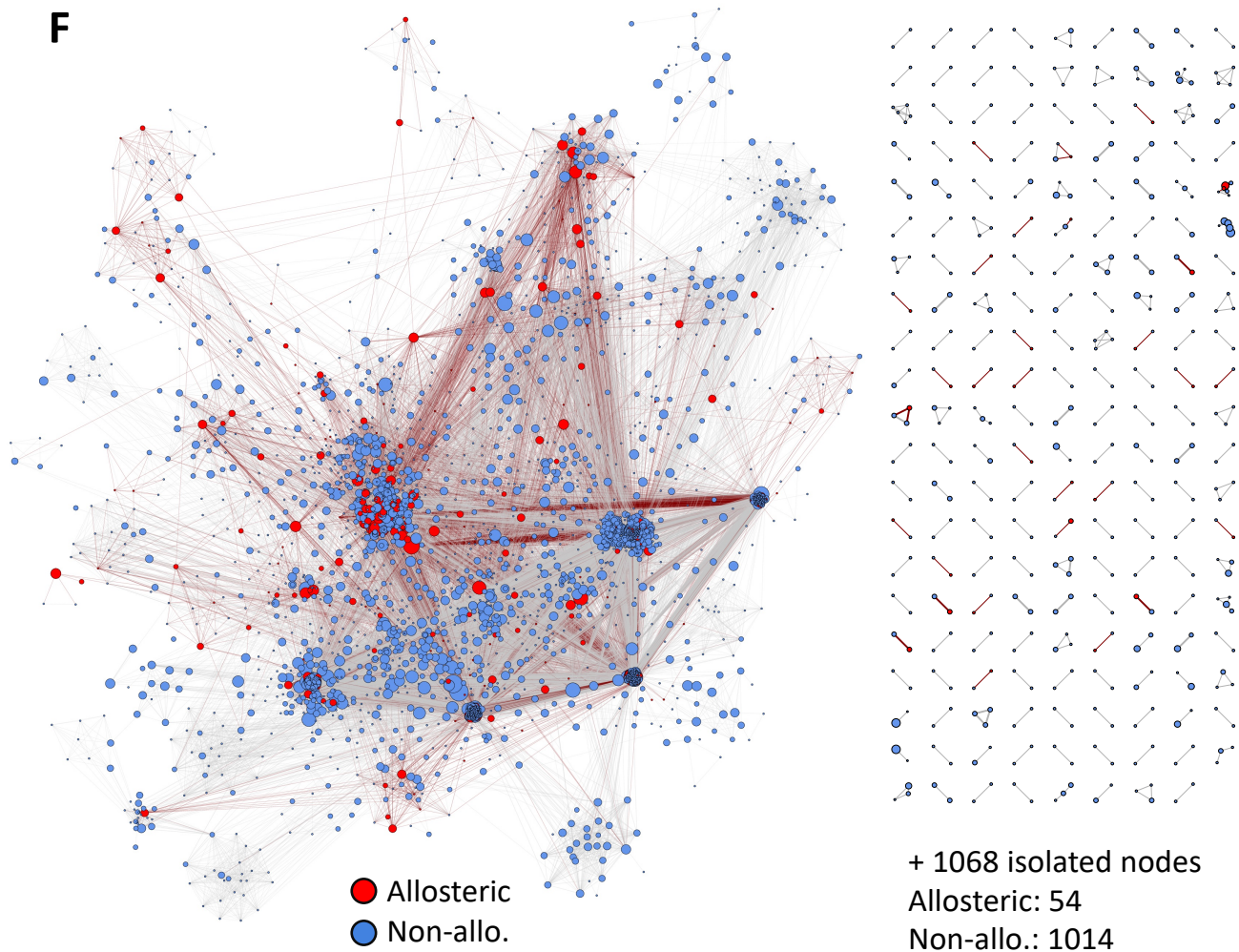

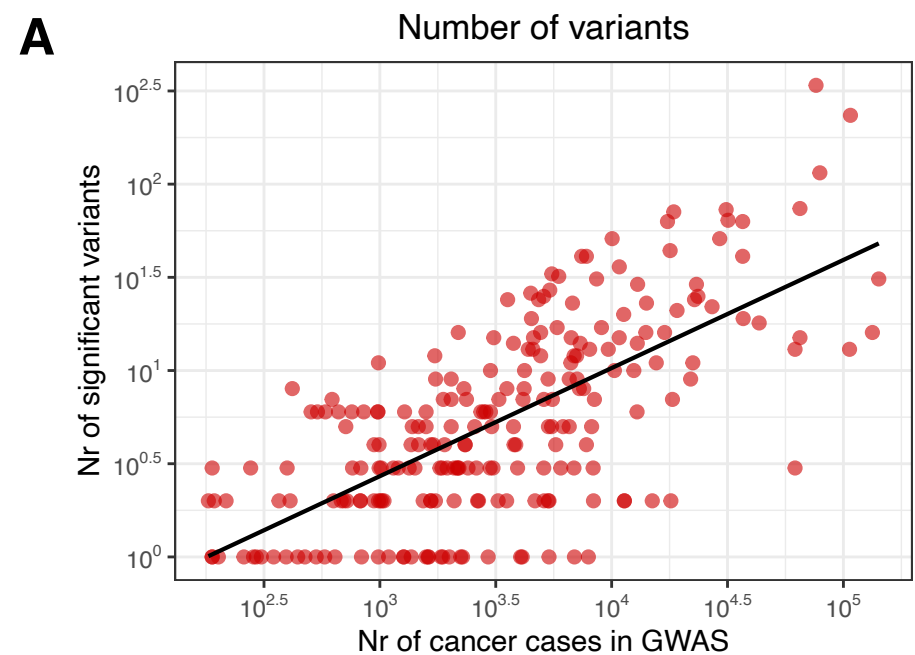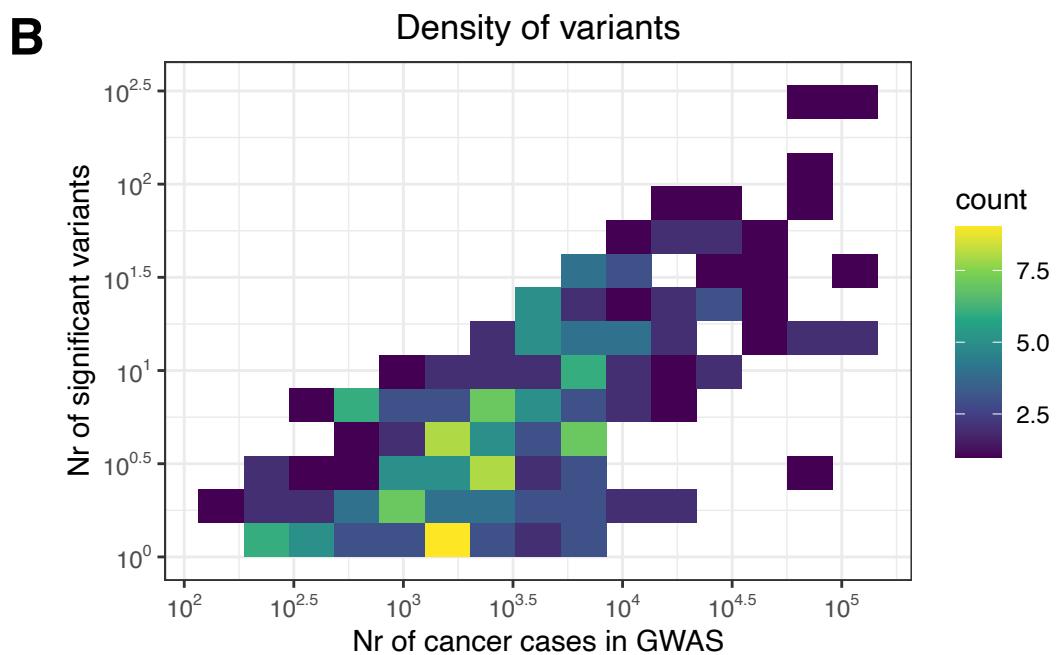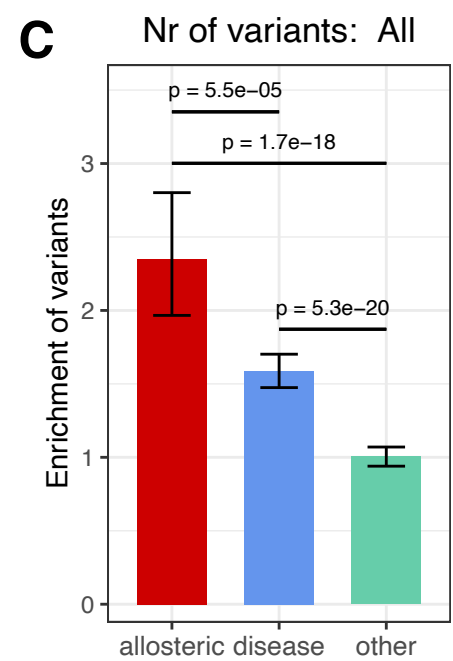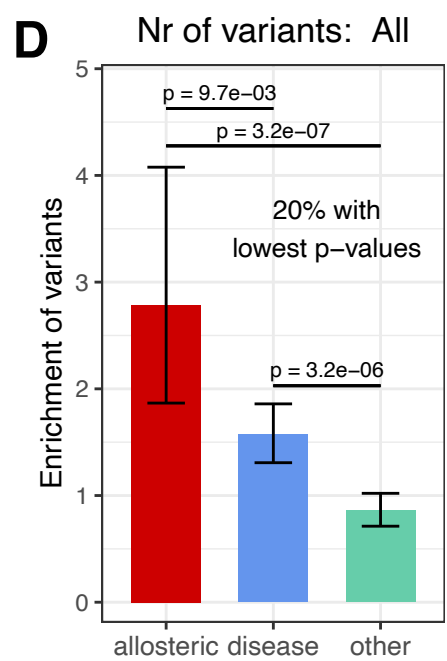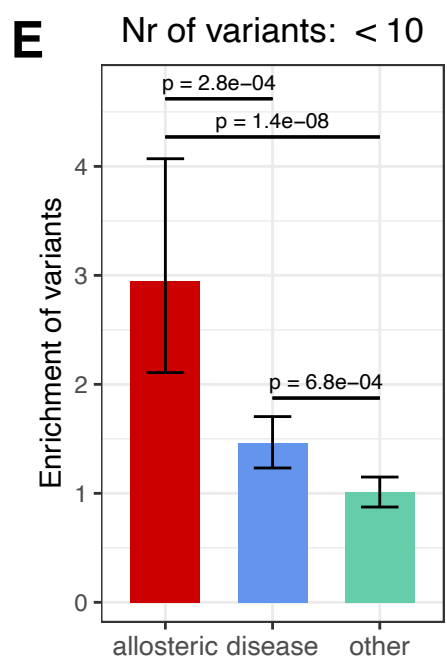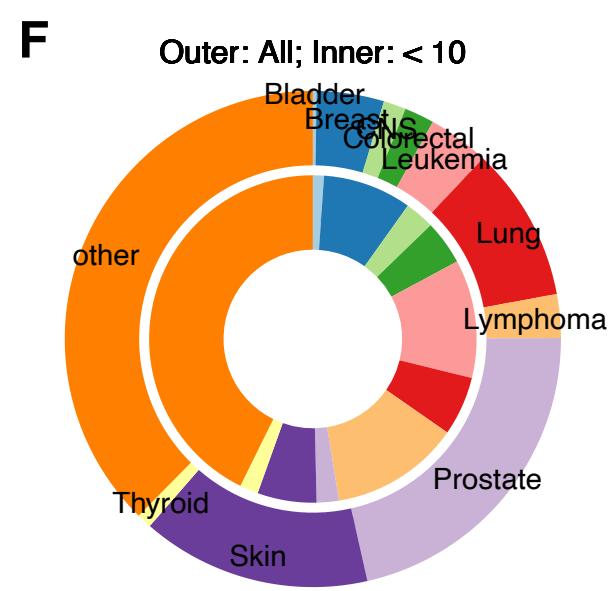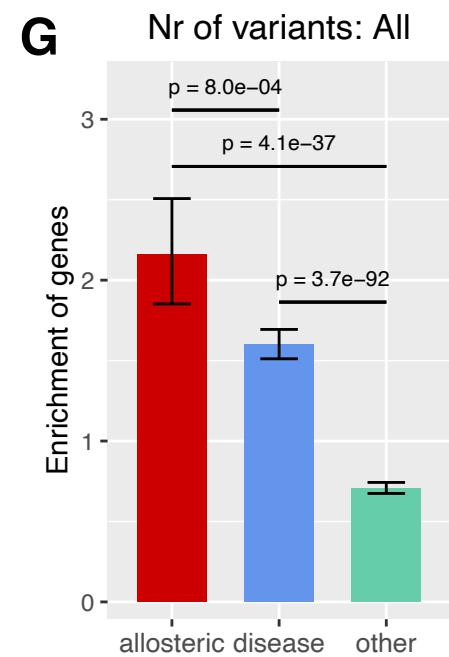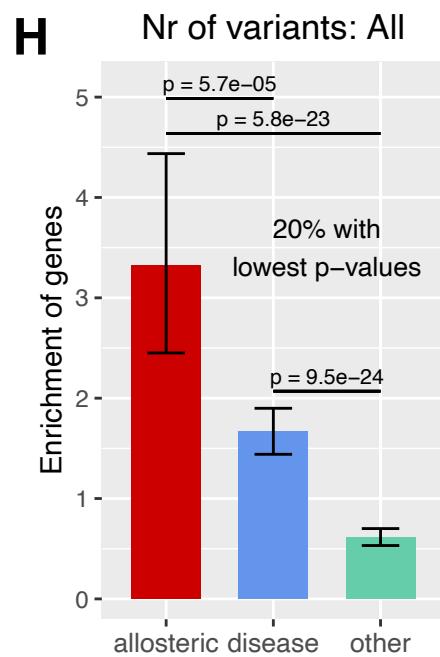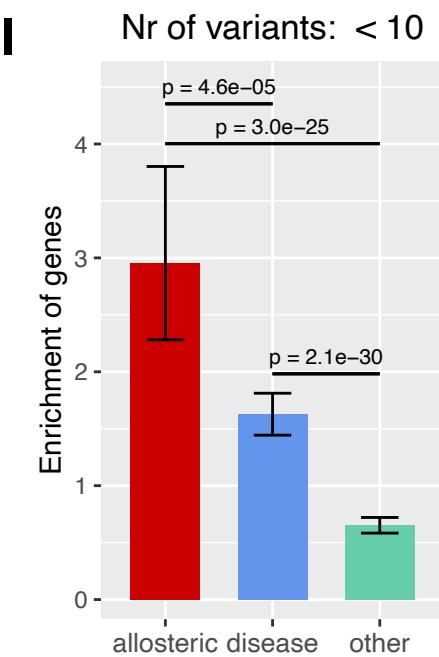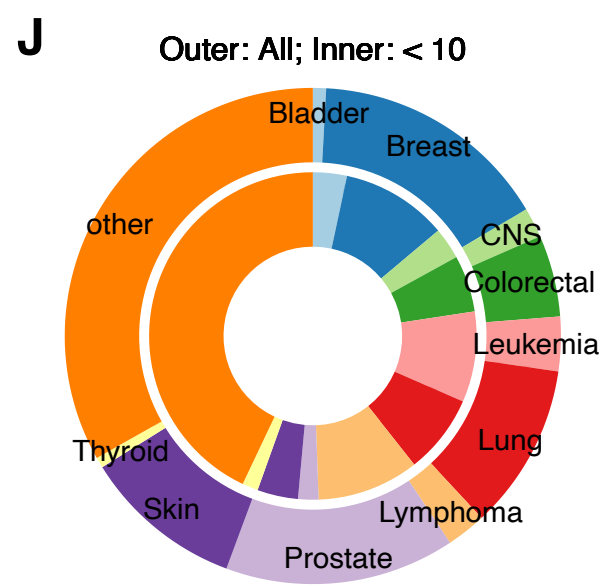

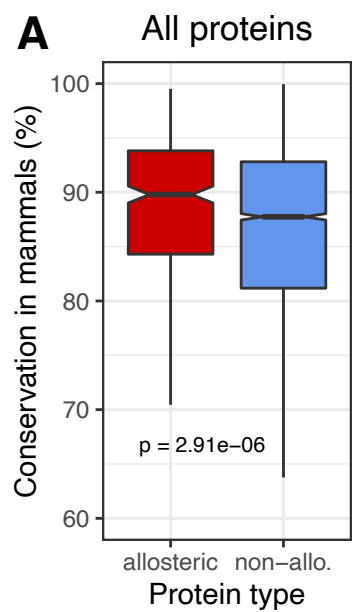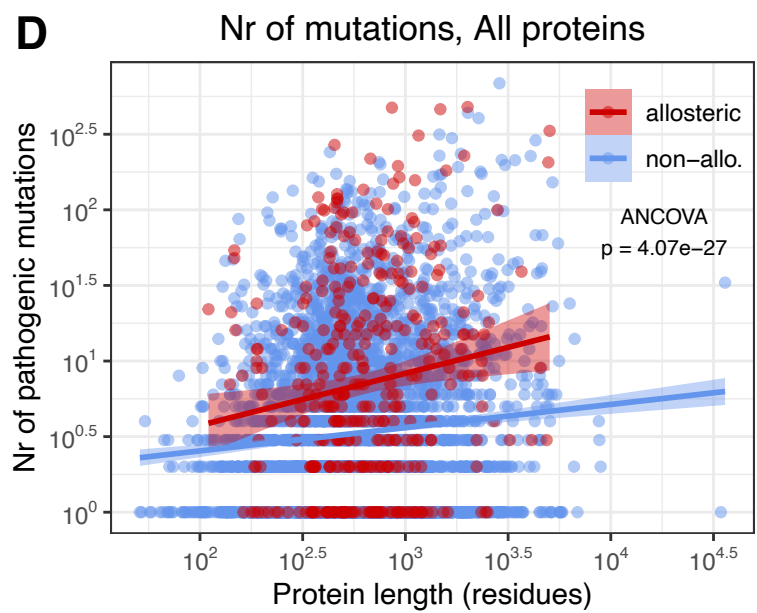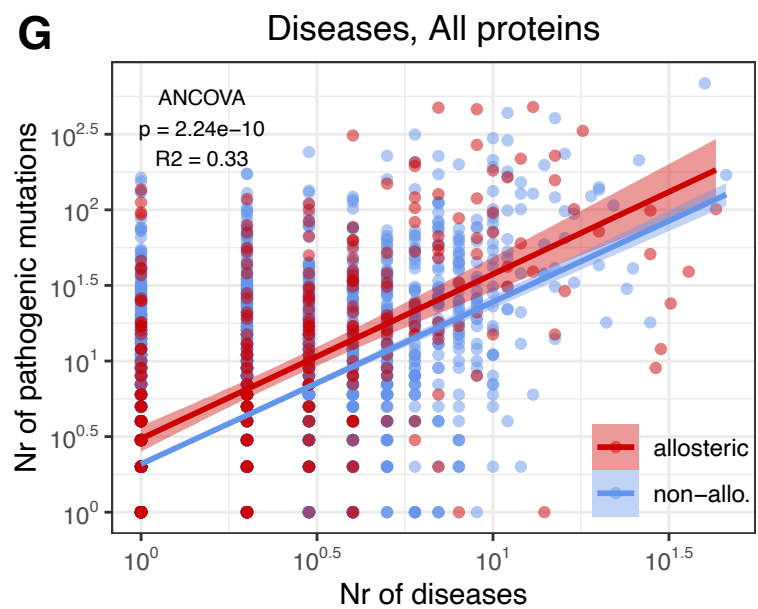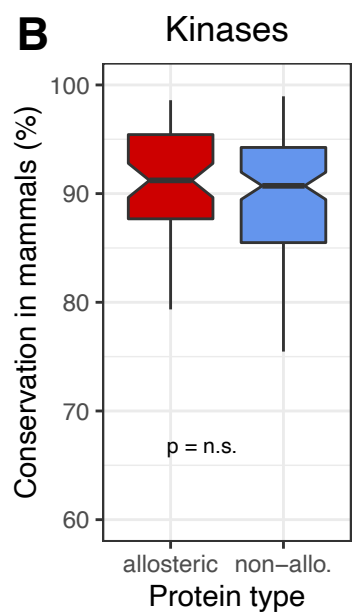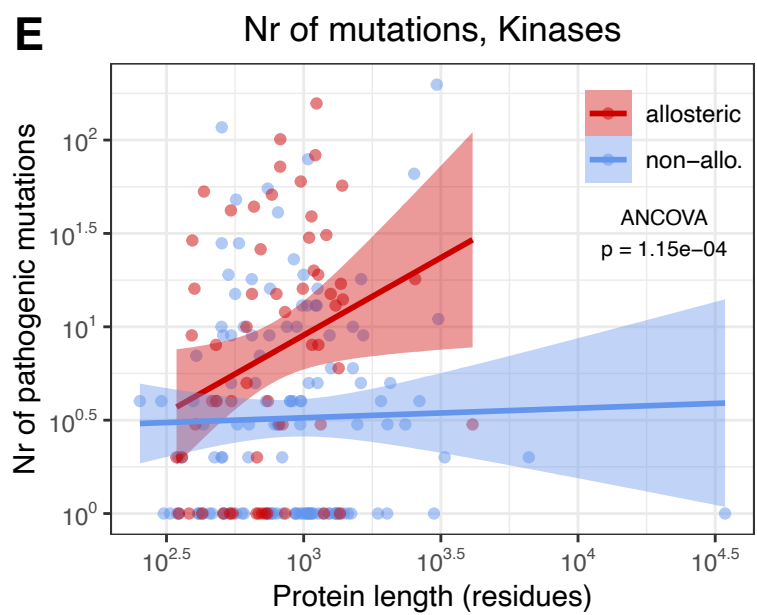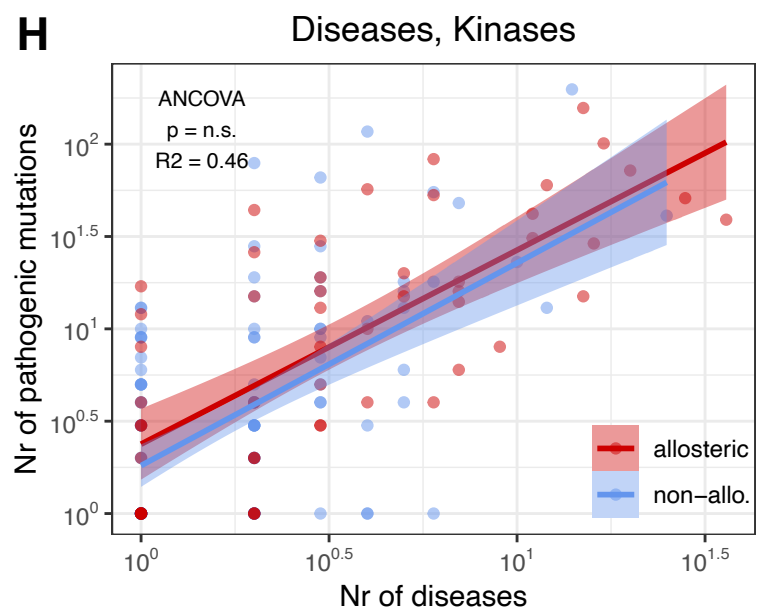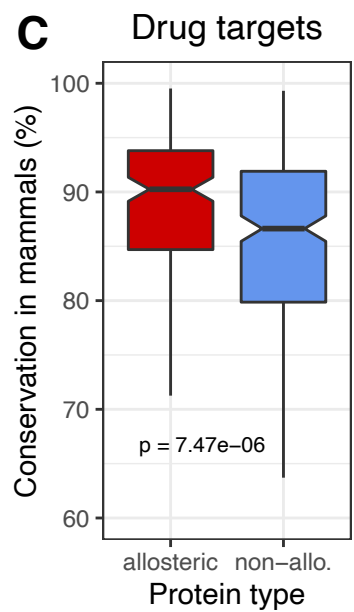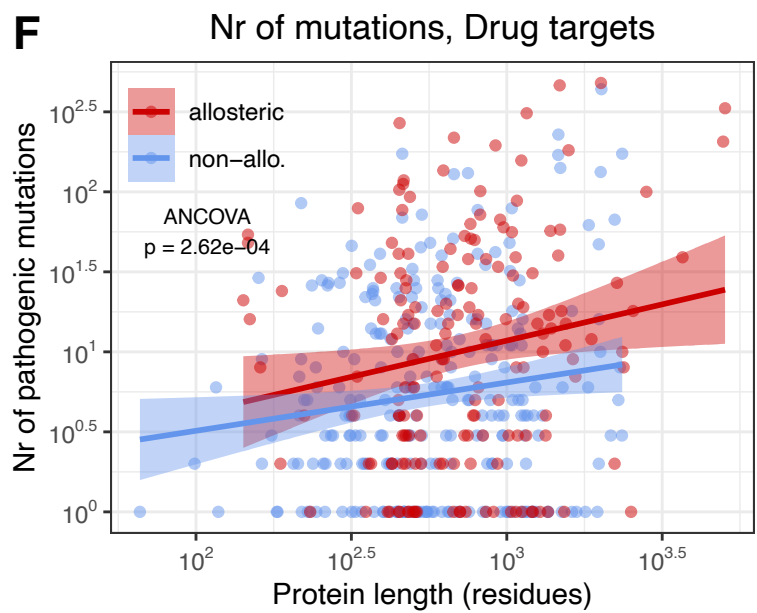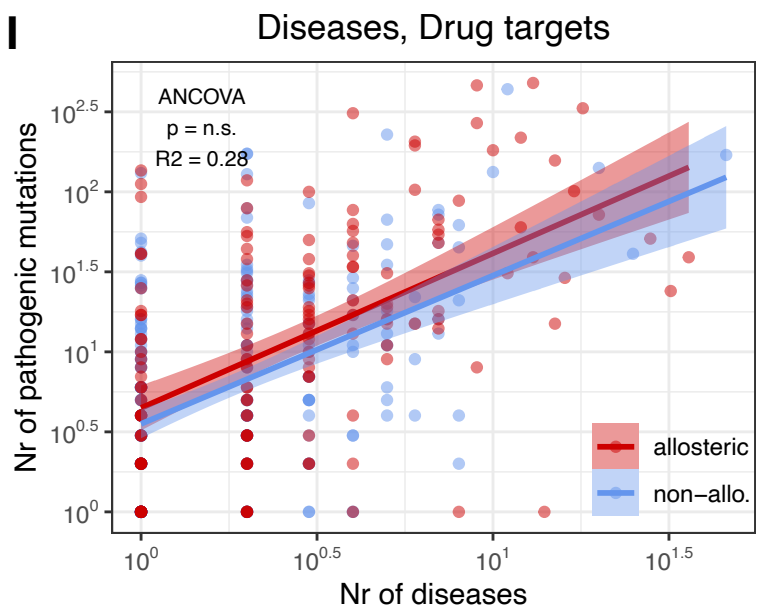

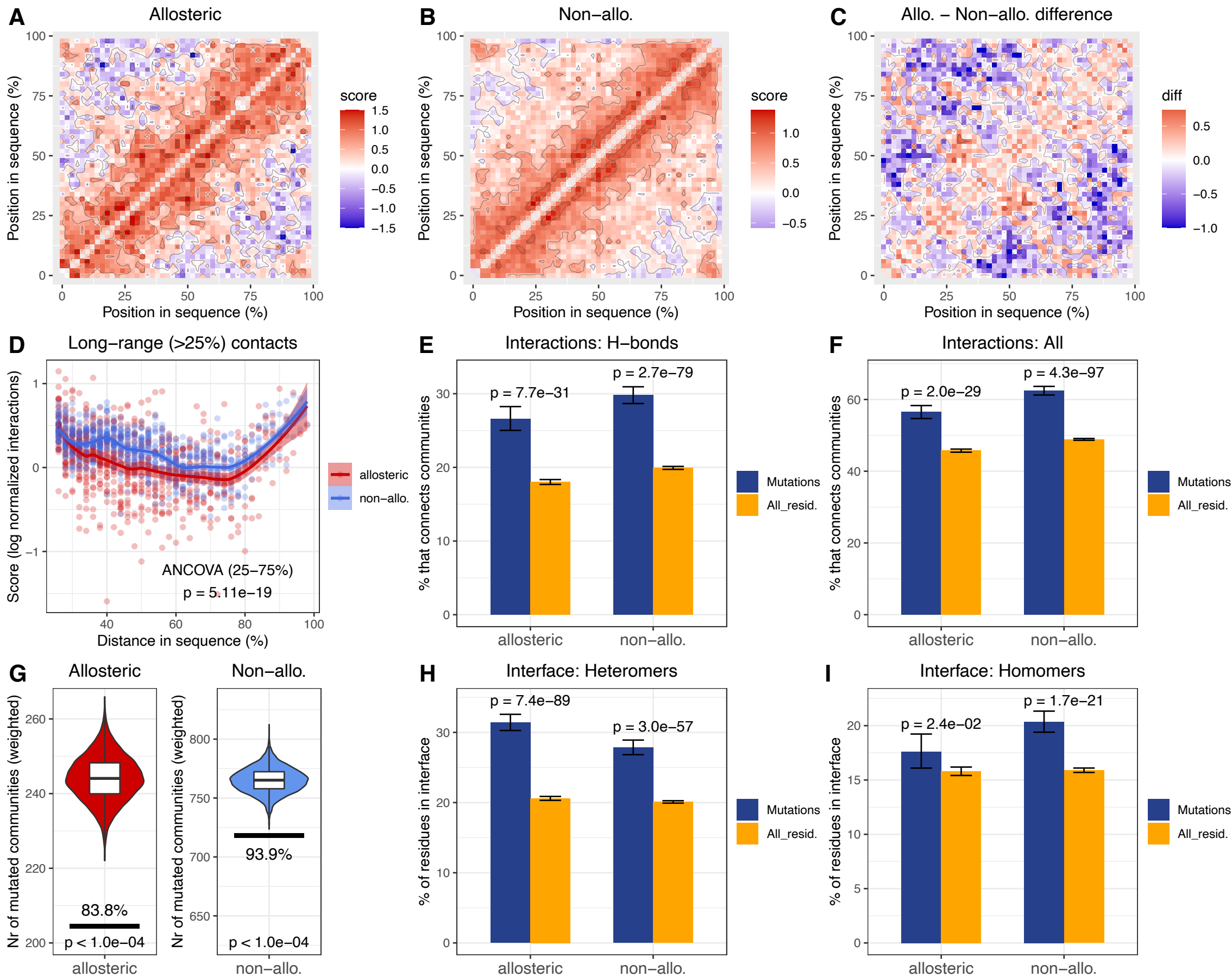

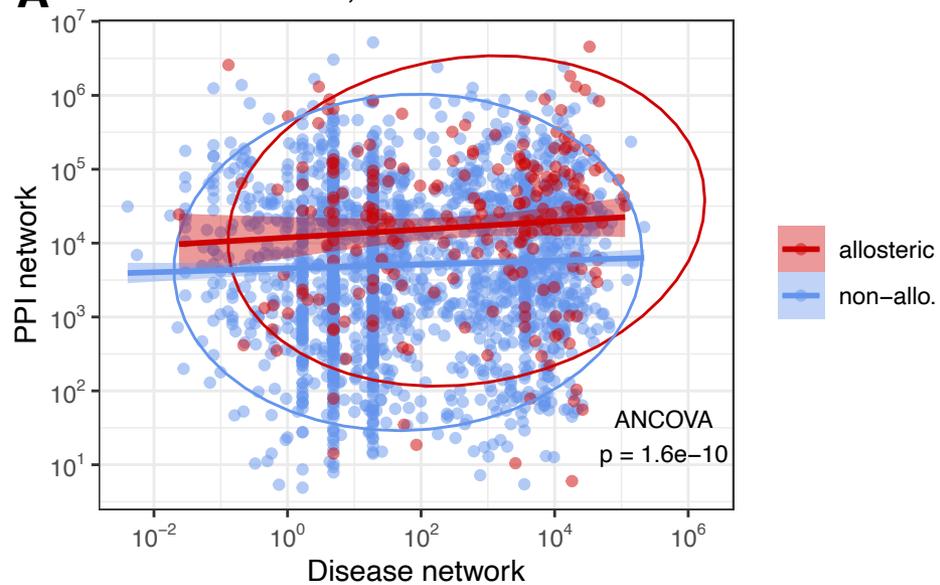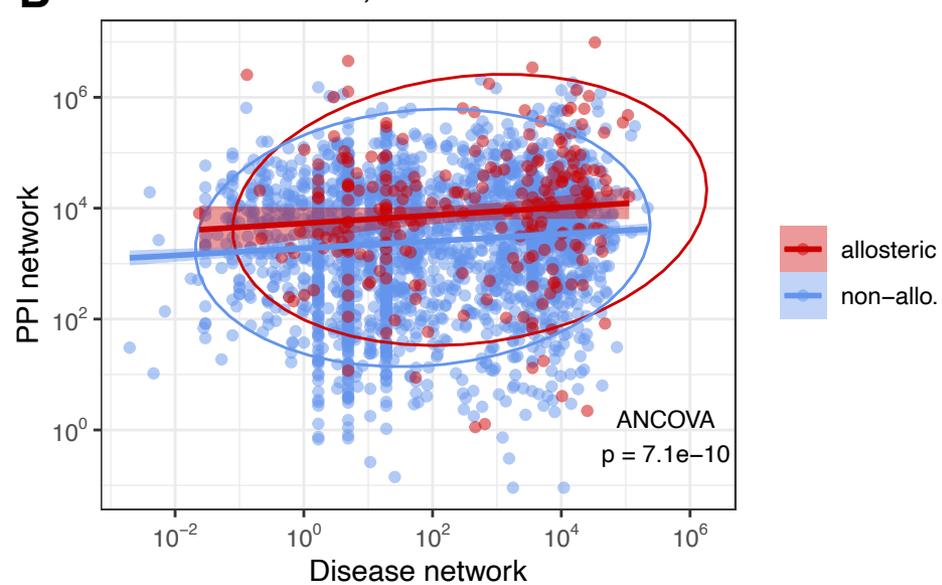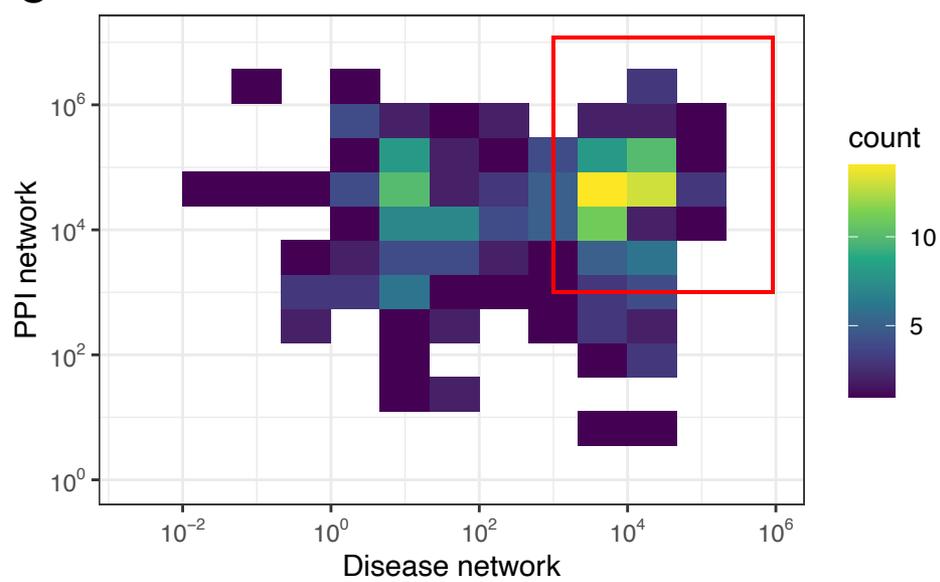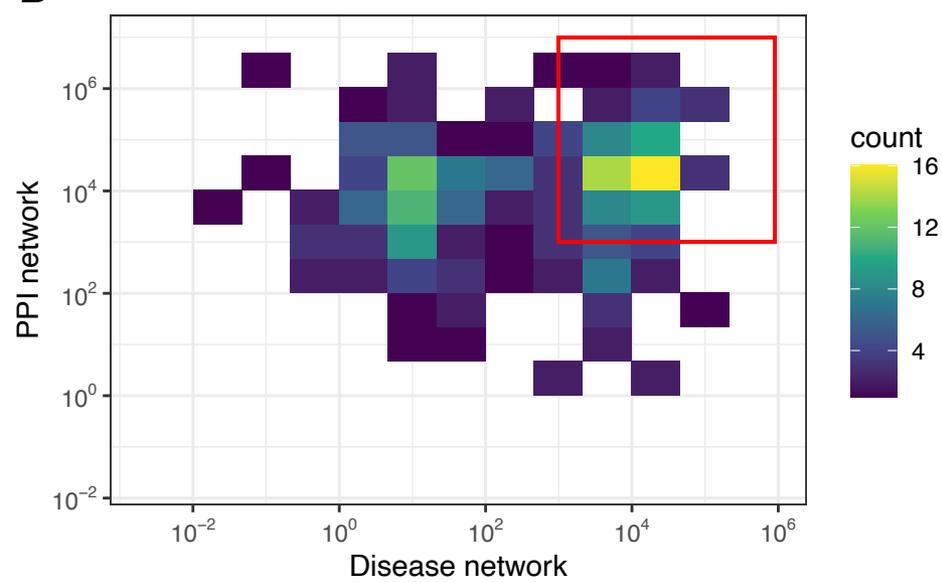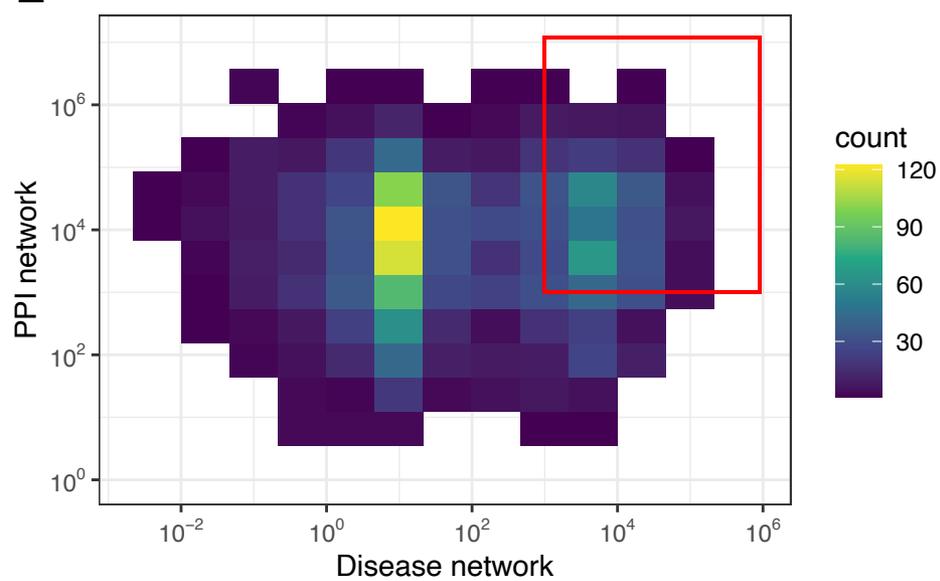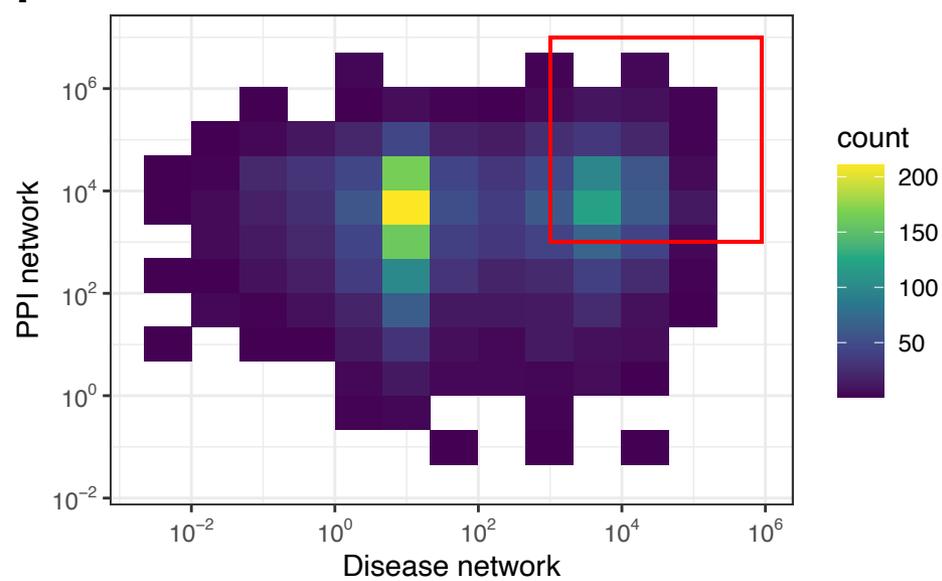